\documentclass[twocolumn,preprintnumbers,amsmath,amssymb, superscriptaddress, prbr]{revtex4}

\usepackage{graphicx}
\usepackage{epstopdf}
\usepackage{dcolumn}
\usepackage{bm}
\usepackage{color}
\usepackage{textcomp}
\definecolor{red}{rgb}{1,0,0}

\definecolor{blue}{rgb}{0,0,1}

\begin{document}
\preprint{APS}

\title{Exchange anisotropy as mechanism for spin-stripe formation in frustrated spin chains}
\author{M. Pregelj}
\affiliation{Jo\v{z}ef Stefan Institute, Jamova 39, 1000 Ljubljana, Slovenia}
\author{O. Zaharko}
\affiliation{Laboratory for Neutron Scattering, PSI, CH-5232 Villigen, Switzerland}
\author{M. Herak}
\affiliation{Institute of Physics, Bijeni\v{c}ka c. 46, HR-10000 Zagreb, Croatia}
\author{M. Gomil\v{s}ek}
\affiliation{Jo\v{z}ef Stefan Institute, Jamova 39, 1000 Ljubljana, Slovenia}
\author{A. Zorko}
\affiliation{Jo\v{z}ef Stefan Institute, Jamova 39, 1000 Ljubljana, Slovenia}
\author{L. C. Chapon}
\affiliation{Institut Laue-Langevin, BP 156, FR-38042 Grenoble Cedex 9, France}
\author{F. Bourdarot}
\affiliation{Institut Laue-Langevin, BP 156, FR-38042 Grenoble Cedex 9, France}
\author{H. Berger}
\affiliation{Ecole Polytechnique F\'{e}d\'{e}rale de Lausanne, CH-1015 Lausanne, Switzerland}
\author{D. Ar\v{c}on}
\affiliation{Jo\v{z}ef Stefan Institute, Jamova 39, 1000 Ljubljana, Slovenia}
\affiliation{Faculty of Mathematics and Physics, University of Ljubljana, Jadranska c. 19, 1000 Ljubljana, Slovenia}
\date{\today}

\begin{abstract}

We investigate the spin-stripe mechanism responsible for the peculiar nanometer modulation of the incommensurate magnetic order that emerges between the vector-chiral and the spin-density-wave phase in the frustrated zigzag spin-1/2 chain compound $\beta$-TeVO$_4$.
A combination of magnetic-torque, neutron-diffraction and spherical-neutron-polarimetry measurements is employed to determine the complex magnetic structures of all three ordered phases.
Based on these results, we develop a simple phenomenological model, which exposes the exchange anisotropy as the key ingredient for the spin-stripe formation in frustrated spin systems.

\end{abstract}

\pacs{}
\maketitle


Textured phases, frequently found in nature, typically develop as a result of conflicting interactions that favor rival ground states.
In biological systems, alternating patterns have been explained by couplings between competing order parameters \cite{Andelman, Seifert, Baumgart}, whereas in strongly correlated electron systems stripe phases have been related to the competition between short- (e.g., exchange) and long-range (e.g., dipolar) interactions, leading, for instance, to stripe domains in feromagnetic films \cite{DeBell, Portmann, Mu, Giuliani} or charge patterns in superconductors \cite{Tranquada, Emery, Vojta, Ghiringhelli, Wu2013}.
Moreover, a theoretical study by Edlund $et$ $al.$ \cite{Edlund} showed that a degeneracy of eigenstates in discretized-spin models allows for stripe formation that does not depend on the details of the microscopic interactions.
Despite the ubiquity of inhomogeneous phases, the recent discovery of a peculiar antiferromagnetic stripe phase -- 
a nanometer-scale modulation of the underlying incommensurate magnetic order in $\beta$-TeVO$_4$ \cite{Pregelj}  -- exceeds the reach of any known model and thus calls for a new, general explanation.

In the monoclinic structure of $\beta$-TeVO$_4$ (space group $P2_1/c$) distorted corner-sharing VO$_5$ pyramids with magnetic V$^{4+}$ ions form zigzag spin-1/2 chains that run along the crystalographic $c$ axis \cite{Meunier, Savina}.
Geometrical frustration stems from the competition between the ferromagnetic nearest-neighbor superexchange interaction $J_1$\,$\sim$\,$-$38\,K and the antiferromagnetic next-nearest-neighboring interaction $J_2$\,$\sim$\,$-$0.8\,$J_1$, while the chains are coupled by more than an order of magnitude weaker, also frustrated, interactions.
As a result, an incommensurate amplitude-modulated, i.e., a spin-density-wave (SDW), state defined by the magnetic wave vector {\bf k}\,=\,($-$0.195,\,0,\,0.413) is established at $T_{N1}$\,=\,4.65\,K, while a vector chiral (VC) spin order develops below $T_{N3}$\,=\,2.28\,K.
Between these two phases, in the temperature range from $T_{N2}$\,=\,3.28\,K to $T_{N3}$, an intriguing spin-stripe phase emerges that is characterized by additional super-satellite reflections appearing in neutron-diffraction profiles at {\bf k}\,$\pm$\,{\bf $\Delta$k} [{\bf $\Delta$k}(2.5\,K)\,=\,($-$0.030,\,0,\,0.021)], which have weak intensities $I$({\bf k}\,$-$\,$\Delta${\bf k})\,$\sim$\,0.1$I$({\bf k}\,$+$\,$\Delta${\bf k})\,$\sim$\,0.01$I$({\bf k}) \cite{Pregelj}.
The resulting stripe order exhibits a remarkable long-scale modulation that alters the main SDW ordering in contrast to other known stripe patterns in magnetic systems.
The negligible long-range dipolar interactions and the realization of an incommensurate, ``nondiscrete'', magnetic order lead to the suggestion that spin stripes arise from a competition between the SDW and VC phases, while the coupling between the two corresponding order parameters was tentatively assigned to the frustrated interchain interactions \cite{Pregelj}.
However, a firm confirmation of this hypothesis and a microscopic explanation of the stripe-forming mechanism, which should also be important for the understanding of the behavior of other strongly correlated electron systems, was still missing.

Here we present an in-depth study of the magnetism in $\beta$-TeVO$_4$, utilizing magnetic-torque, neutron-diffraction and spherical-neutron-polarimetry measurements, which reveals the details of the long-range order in all three magnetically ordered phases. 
Our results provide new insight into the magnetism of frustrated spin chains allowing us to affirmatively answer the hypothesis about spin-stripe formation in $\beta$-TeVO$_4$ and to complement it by a phenomenological description of the corresponding microscopic mechanism based on exchange anisotropy.
As this type of anisotropy is very common in spin systems, we predict that similar spin-stripe phases should appear in many frustrated spin-chain compounds.

\begin{figure}[!]
\centering
\includegraphics[width=\columnwidth]{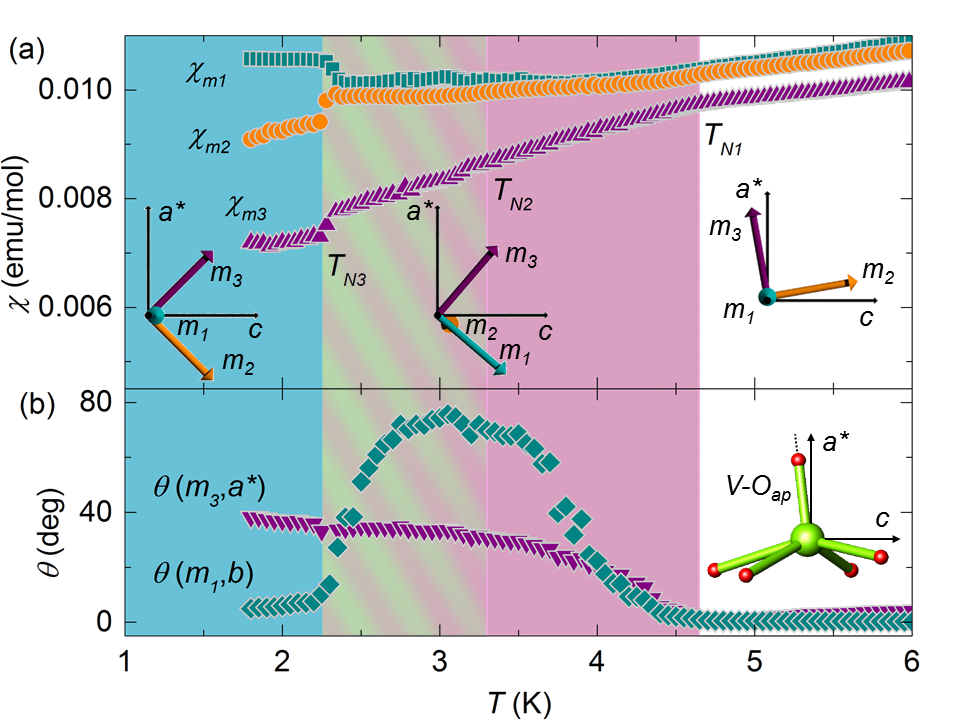}
\caption{The temperature dependence of (a) the eigenvalues of the magnetic susceptibility and (b) of the orientations of the corresponding magnetic eigenaxes $m_1$, $m_2$ and $m_3$ with respect to the crystallographic axes. Insets: (a) a sketch of magnetic eigenaxes $m_1$, $m_2$ and $m_3$ with respect to crystallographic axes, and (b) VO$_5$ pyramid.}
\label{fig:torque}
\end{figure}

We begin by magnetic-torque measurements that are especially useful for probing macroscopic changes of the magnetic symmetry that occur at magnetic phase transitions.
The measured torque is directly proportional to the anisotropy in the magnetic susceptibility $\Delta \chi_{xy}=\chi_x - \chi_y$ in the measurement ($xy$) plane.
Hence, its angular dependence reveals the directions of maximal and minimal susceptibility within this plane.
We measured the angular dependencies of magnetic torque in $\beta$-TeVO$_4$ in magnetic fields up to 0.8\,T applied perpendicularly to all three crystallographic planes between 300 and 1.5 K.
This data disclose the temperature dependence of the full magnetic-susceptibility tensor, i.e., its eigenvalues $\chi_{mi}$ and eigenaxes {\bf m}$_i$, $i$\,=\,1-3 (Fig.\,\ref{fig:torque}) \cite{supp}.

In the paramagnetic phase, i.e., above $T_{N1}$, we find a pronounced anisotropy $\chi_a<\chi_c<\chi_b$, in agreement with previous magnetic susceptibility measurements \cite{Savina}.
However, in contrast to those results, we find that the difference between $\chi_b$ and $\chi_c$ is much smaller than between $\chi_a$ and $\chi_c$.
Our result is consistent with the crystal structure, as the orientation of the VO$_5$ pyramids [inset in  Fig.\,\ref{fig:torque}(b)] implies an easy-plane-like ($bc$) anisotropy.
In fact, the anisotropy eigenaxes within the $ac$ plane coincide with the direction of the bond between the vanadium and the apical oxygen, V-O$_{\textup{ap}}$ [insets in  Fig.\,\ref{fig:torque}], and the direction perpendicular to it, as they depart $\sim\,$10$^\circ$ from the $a$ and $c$ axes, whereas the $b$ axis is an exact eigenaxis in agreement with the 2$_y$ crystal symmetry.

At low temperatures, the direction of the eigenaxes and the size of the magnetic susceptibility change due to the establishment of long-range magnetic order (Fig.\,\ref{fig:torque}). 
At $T_{N1}$, the changes are rather subtle, implying that the magnetic susceptibility is still dominated by a paramagnetic response.
This is consistent with an SDW-type of order \cite{Pregelj}, where a significant part (more than half) of the vanadium magnetic moments is still fluctuating.
As temperature approaches $T_{N2}$, the magnetic-susceptibility eigenaxes rotate [Fig.\,\ref{fig:torque}(b)] breaking the 2$_y$ symmetry, yet the eigenvalues remain basically unchanged [Fig.\,\ref{fig:torque}(a)]. 
This indicates that the size of the ordered magnetic moments does not change notably.

Upon approaching $T_{N3}$, the magnetic-susceptibility eigenaxes rotate again, but this time also the eigenvalues change appreciably (Fig.\,\ref{fig:torque}), in agreement with a structural deformation occurring at $T_{N3}$ \cite{Weickert}.
Namely, {\bf m}$_1$ aligns back along the $b$ axis and $\chi_{m1}$ becomes much larger than $\chi_{m2}$.
The latter is slightly reduced and the related eigenaxis now points almost along ($-$1\,0\,1), while $\chi_{m3}$, with {\bf m}$_3$ pointing roughly along (1\,0\,1), is also reduced and remains substantially smaller than $\chi_{m2}$.
Clearly, the easy-plane symmetry, i.e., $\chi_{m1}$ being similar to $\chi_{m2}$, is broken at $T_{N3}$ when the VC phase is established.
This stands in contrast to the response found in archetypal chiral, i.e., screw-type, phases where the magnetic moments are coplanar \cite{Bursill, Arima, Tokura}, which should manifest in a uniaxial susceptibility tensor. 
The response in the VC phase thus suggests a more complicated magnetic order with broken uniaxial symmetry.

To obtain detailed information about the magnetic orders, we performed neutron diffraction experiments on a high-quality single crystal \cite{Pregelj} of a size $2\times 3 \times 4$\,mm$^3$.
Intensities of 158 magnetic reflections were measured at 3.5 and 1.7\,K, i.e., in the SDW and VC phases, respectively, on the TriCS diffractometer at the Paul Scherrer Institute, Switzerland.
Complementary spherical-neutron-polarimetry experiments were performed at the Institute Laue Langevin, Grenoble, France on D3 and IN22 instruments equipped with CRYOPAD devices. 
Polarimetry matrices were measured for several strongest magnetic reflections for two different scattering planes in all magnetically ordered phases, i.e., at 3.5, 2.5 and 1.7\,K.

\begin{figure} [!]
\includegraphics[width=0.48\textwidth]{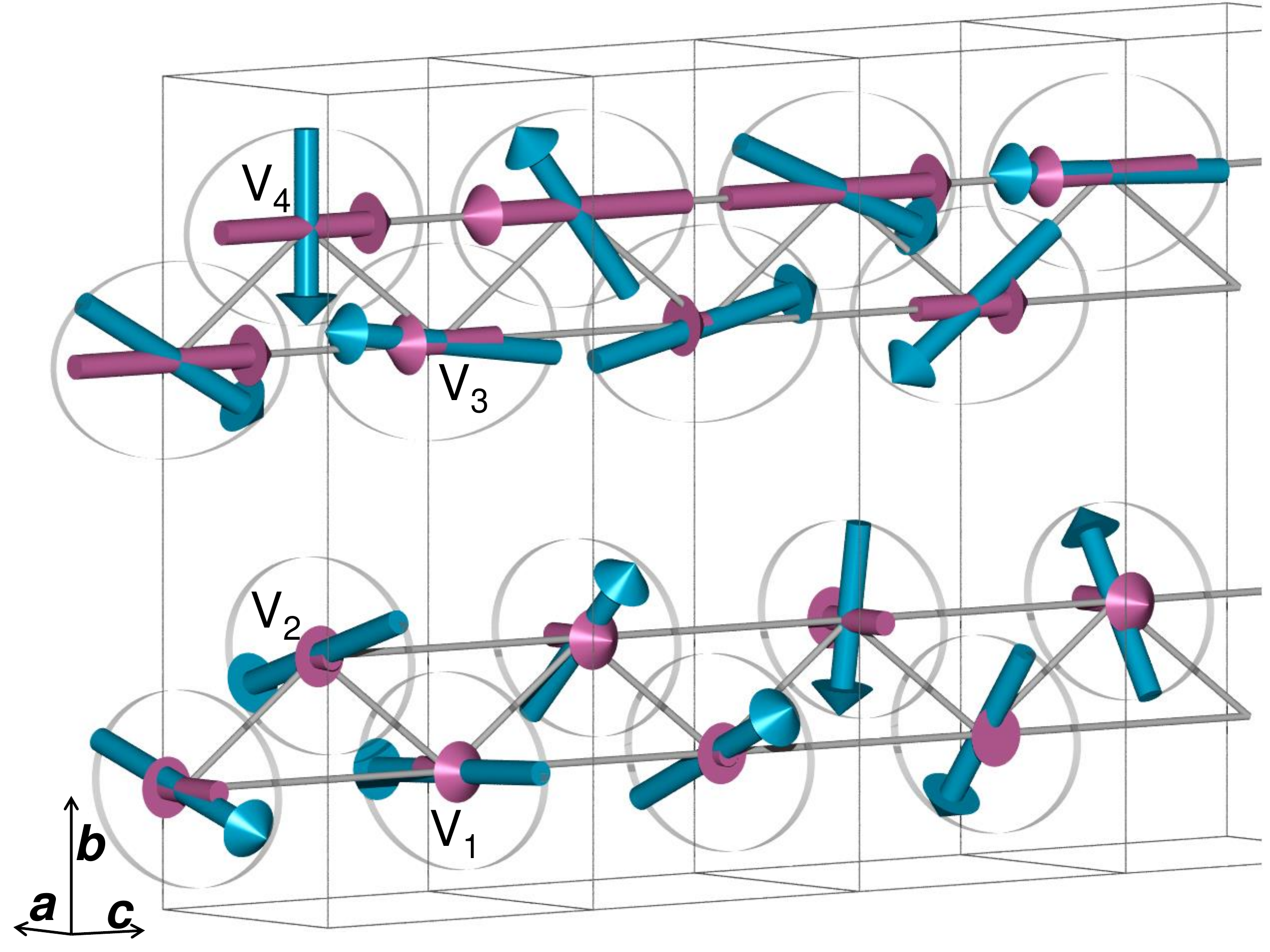}
\caption{Magnetic structures in the SDW (magenta) and VC (cyan) phases derived from the combined refinement of the magnetic peak intensities and the polarization matrices.}
\label{mag-struc}
\end{figure}

An important first step prior to the refinement of the magnetic structure is representation analysis, which allows postulating symmetry restrictions based on the crystal lattice and the magnetic wave vector.
In particular, the magnetic wave vector {\bf k}\,$\sim$\,($-0.2,\,0,\,0.42$) \cite{Pregelj}
means that the only symmetry element left in the magnetically ordered phase is the $c(x,1/4,z)$ glide plane, which is orthogonal to the $b$ axis.
Hence, the simplest magnetic structure model consists of one of two possible irreducible representations, $\Gamma_1$ and $\Gamma_2$ (Table\,\ref{irreps}).
These relate the two vanadium sites within each of the two crystallographically equivalent spin chains, i.e., V$_1$ with V$_2$ and V$_3$ with V$_4$ (Fig.\,\ref{mag-struc}).%
\begin{table}[tb]
	\centering
	\caption{Irreducible representations $\Gamma_1$ and $\Gamma_2$ of the little group for the magnetic wave vector {\bf k}\,$\sim$\,($-0.2, 0, 0.42$) in the space group $P2_1/c$. $k_c$ denotes the $c$ component of {\bf k}.}
		\begin{tabular*}{\linewidth}{@{\extracolsep{\fill}} c c c }
   \hline \hline
   	 V-site& $\Gamma_1$ & $\Gamma_2$\\
   	 \hline
   	 $x$, $y$, $z$ & $(u, v, w)$ & $(u, v, w)$\\
   	 $x$, $-y+1/2$, $z+1/2$ & $(u, -v, w)e^{-i\pi k_c}$		 & $(-u, v, -w)e^{-i\pi k_c}$\\
   	 \hline \hline
		\end{tabular*}
			\label{irreps}
\end{table}
In addition, as the magnetic order is dictated by the frustrated interactions $J_1$ and $J_2$, we expect that magnetic moments on the individual chain form a coplanar spiral \cite{Bursill} that may be preceded by a collinear SDW-type order \cite{Haaris2007}.
Since $\Gamma_1$ ($\Gamma_2$) imposes opposite signs of the $b$ ($a$ and $c$) components of the magnetic moments at the neighboring V$_i$ and V$_{i+1}$, $i$\,=\,1,3 sites (Table\,\ref{irreps}), i.e., contradicting the spiral and SDW orders, we confine $\Gamma_1$ to the $a$ and $c$ components and $\Gamma_2$ to the $b$ component.

The combined refinements \cite{FTOB-PSD} of the integrated magnetic-peak intensities and polarization matrices for the SDW and VC phases yield rather unusual results \cite{supp}. 
In the SDW phase the derived magnetic structure model corresponds to the $\Gamma_1$ representation with magnetic moments pointing along the $c$ axis on one chain and along the $a$ axis on the neighboring chain (Fig.\,\ref{mag-struc}).
This indicates the importance of the frustrated interchain interactions that can also have a finite antisymmetric-exchange, i.e., Dzyaloshinskii-Moriya, contribution \cite{Weickert}. 
An orthogonal spin arrangement on neighboring chains has been predicted for a spatially anisotropic triangular Heisenberg lattice when highly frustrated interchain interactions are much smaller than the intrachain ones \cite{Weichselbaum}, which is indeed the case for $\beta$-TeVO$_4$. 
Moreover, at 3.5\,K, the derived amplitude of the magnetic moments on the first chain is 0.7(1)\,$\mu_B$, while it amounts to only 0.4(1)\,$\mu_B$ on the second one, significantly less than 1\,$\mu_B$ expected for a full magnetic moment of the V$^{4+}$ $S$\,=\,1/2 ion.
Obviously, a major part of the magnetic moments is not yet developed, i.e., a feature characteristic of frustrated spin systems, corroborating the magnetic-torque results. 
Finally, the fact that the $b$ component of the magnetic moments does not develop implies that $J_1$ and $J_2$ are anisotropic and are weaker for this spin component.

\begin{table} [!]
\caption{Polarization matrix terms $P_{\alpha\beta}$, where $\alpha,\beta$\,=\,$x,y,z$, at 2.5\,K at the main (left) and satellite (right) reflections for the ($-hk2h$) scattering plane.
\label{SNPstripe}}
\begin{ruledtabular}
\begin{tabular}{ccccccc}
 &\multicolumn{3}{c}{main} & \multicolumn{3}{c}{satellite} \\
\hline
      &  $x$    &  $y$      &   $z$   &   $x$    &   $y$     &   $z$\\
$x$ &-0.94(4) & -0.05(4) & 0.08(4) & -1.2(2) & -0.0(1) & 0.0(1)\\
$y$ & 0.03(4) & -0.82(4) & 0.04(4) & 0.0(1)  & 0.9(1) & -0.1(1)\\
$z$ & 0.01(4) &  0.00(4) & 0.83(4) & 0.0(1) & -0.0(1) & -1.3(2)\\
\end{tabular}
\end{ruledtabular}
\end{table}

In the VC phase, elliptical spin spirals with the aspect ratio of 1:0.70(5) (at 1.7\,K) develop on both chains breaking the $c(x,1/4,z)$ symmetry. 
The inequivalence of the two chains is retained, as the corresponding spin spirals lie in almost perpendicular planes, i.e., $ab$ and $bc$ (see Fig.\,\ref{mag-struc}), which is consistent with the non-uniaxial form of the magnetic susceptibility tensor found below $T_{N3}$ (Fig.\,\ref{fig:torque}).
In addition, the two spirals have also different orientations of their major axes, one pointing along the $b$ axis and the other one along the $c$ axis. 
Our results thus clearly show that the magnetic structure in the VC phase is related with the establishment of the $b$ magnetic component associated with the $\Gamma_2$ representation, in agreement with Ref.\,\onlinecite{Pregelj}.

In the stripe phase, at 2.5\,K, the measured polarization matrix at the strongest magnetic reflection {\bf k}\,$\sim$\,($-0.2,\,0,\,0.42$) is very similar to the one measured in the SDW phase (at 3.5\,K).
This shows that the dominant magnetic order in the two phases is approximately the same.
In contrast, at the weak satellite reflection {\bf k}\,$+$\,{\bf $\Delta$k}, the $P_{yy}$ and $P_{zz}$ matrix elements change signs (Table\,\,\ref{SNPstripe}), indicating that the corresponding magnetic ordering is restricted to the ($-hk2h$) plane and is thus almost perpendicular to the dominant one. 
Still, the chiral matrix terms, $P_{xy}$ and $P_{xz}$, are very small for both reflections, signifying that in the stripe phase both magnetic modulations are associated solely with the amplitude variation of the magnetic moments, i.e., are of the SDW type.
This reveals that the nanoscale stripe modulation emerges when the main magnetic order inherited from the SDW phase is accompanied with an additional SDW modulation with an orthogonal alignment of the magnetic moments and a slightly different periodicity.

\begin{figure}[!]
\centering
\includegraphics[width=\columnwidth]{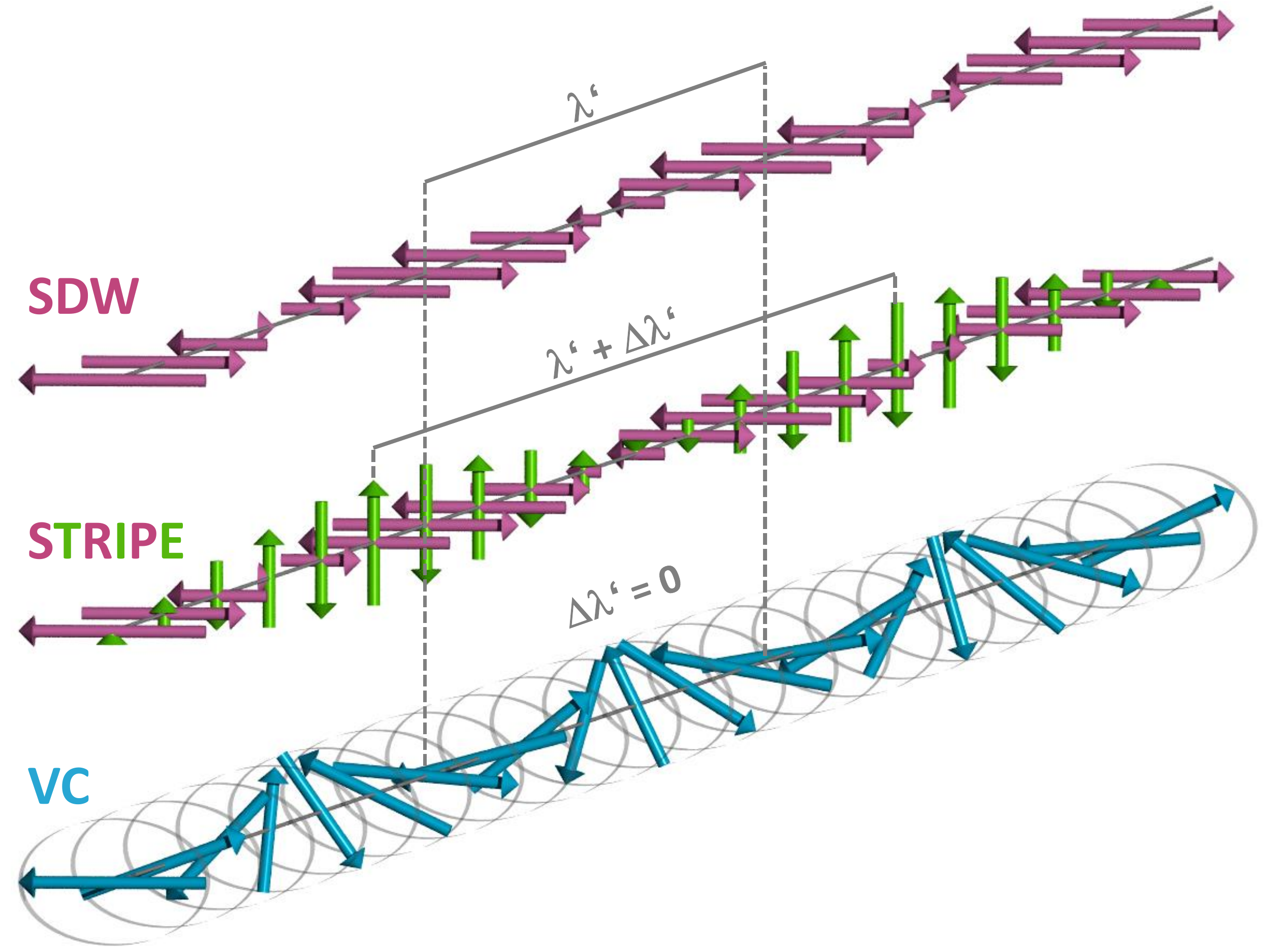}
\caption{Magnetic order at the V$_1$ site along the $c$ axis derived for the SDW (top) and VC (bottom) phases, complemented with a reconstruction of the order in the stripe phase (middle). $\lambda$' corresponds to the deviation of {\bf k} from 1/2, i.e., from the antiferromagnetic modulation, while $\Delta\lambda$' corresponds to {\bf $\Delta$k}.}
\label{fig:phenomenological}
\end{figure}

Now, we suppose that the magnetic order in the VC phase can be decomposed into two SDW components; one which already exists in the SDW phase and corresponds to magnetic moments in the $ac$ plane and the other that corresponds to the magnetic moments aligned along the $b$ axis.
Such a scenario fully agrees with the constraints derived for the SDW associated with the {\bf k}\,+\,{\bf $\Delta$k} reflections, which develop in the stripe phase and thus probably correspond to the $\Gamma_2$ representation.
In other words, the spiral magnetic order in the VC phase is most likely a superposition of the two SDW orders that form stripes in the preceding phase  (Fig.\,\ref{fig:phenomenological}), one corresponding to the $\Gamma_1$ and the other to the $\Gamma_2$ representation.
The condition for the transition into the VC phase is met when the two magnetic vectors associated with the two SDW orders become equal, i.e, when {\bf $\Delta$k} becomes zero.

The above experimental findings allow us to develop a phenomenological description of the stripe-forming mechanism in terms of a free energy model.
A minimal free energy model of multiple coexisting modulation wave vectors has been written in Refs. \onlinecite{Shapiro} and \onlinecite{Forgan}.
The general constraints are imposed by the fact that the free energy is a scalar and is thus invariant under the time inversion and the symmetry operations that define the crystal structure.
In particular, the former demands that terms in the free energy always involve an even number of sublattice magnetizations {\bf M$_{k_i}$}, while the latter requires that the corresponding wave vectors {\bf k}$_i$ sum up into a reciprocal-lattice vector {\bf G} or zero, i.e., $\sum_i${\bf k}$_i$\,=\,{\bf G} \cite{Forgan, Shapiro}.
In the stripe phase of $\beta$-TeVO$_4$, the magnetic structure is defined by three magnetic wave vectors \cite{Pregelj}, i.e., {\bf k}, {\bf k}\,$+$\,{\bf $\Delta$k} and {\bf k}\,$-$\,{\bf $\Delta$k}. 
Consequently, the lowest-order coupling term in the free energy must have the form 
\begin{equation}
\label{MF1}
F \propto M_{\bf -k}^2 M_{{\bf k}+{\bf \Delta k}} M_{{\bf k}-{\bf \Delta k}}.
\end{equation}

Next, we take into the account the fact that spin-stripe formation in an antiferromagnet with negligible long-range dipolar interactions, requires two coupled order parameters.
These were initially associated with the VC and SDW states \cite{Pregelj}.
The 
 temperature evolution of the 
magnetic structure reveals important similarities with magnetoelectric multiferroics \cite{Wang}
, where incommensurate magnetic phases also develop in sequence. 
This 
allows for a general description of spin-stripe formation.
Namely, in analogy to multiferroics \cite{Lawes, Kenzelmann, Haaris} we can assume that the magnetic component {\bf M$_{i}$} corresponding to a given representation $\Gamma_i$ is related to its own magnetic order parameter, i.e., {\bf M$_{\bf k}$}  ($\Gamma_1$) to $\sigma_1$\,=\,$|\sigma_1|e^{i\phi_1}$ and {\bf M$_{{\bf k}\,\pm\,{\bf \Delta k}}$}  ($\Gamma_2$) to $\sigma_2$\,=\,$|\sigma_2|e^{i\phi_2}$ \cite{supp}.
As a result, Eq.\,(\ref{MF1}) transforms into
\begin{equation}
\label{MF3}
F \propto |\sigma_1|^2 |\sigma_2|^2 \text{cos}[2(\phi_1+\phi_2)].
\end{equation}
Setting the arbitrary phase of the first order parameter to zero, we find that the coupling is strongest when the phase for the second order parameter is $\phi_2$\,=\,$\pi/2$.
The latter reflects the orthogonal alignment of the two coexisting SDW components in the stripe phase.
In addition, fourth-order terms in the free energy favor magnetic moments of the same size, that is
fixed-length spins \cite{Haaris2007,Haaris2006}, which is realized below $T_{N3}$, where {\bf $\Delta$k}\,=\,0 and the two SDW modulations merge into a spin-spiral with almost fully developed magnetic moments (Fig.\,\ref{fig:phenomenological}).

The two magnetic order parameters in $\beta$-TeVO$_4$ are clearly related to the two superposed orthogonal SDW components, corresponding to the $\Gamma_1$ and $\Gamma_2$ representations.
Yet, 
in contrast to the existing studies of multiferroic materials \cite{Lawes, Kenzelmann, Haaris, Wang}, we find that 
these two exhibit different modulations and temperature dependencies, which can occur only in the presence of the magnetic anisotropy.
The anisotropy is indeed sizable, as the room-temperature ($k_BT$\,$\gg$\,$J_1$,$J_2$) magnetic susceptibility shows an anisotropy of $\sim$10\% \cite{Savina}, which is large, but not completely unexpected for a V$^{4+}$ $S$\,=\,1/2 system \cite{Eremin}.
Different modulations of the spin components may thus be associated with the anisotropy of the main exchange interactions, i.e., $J_i$ ($i$\,=\,1,2) for spin components along the $b$ axis ($J_i^{b}$) may differ from its value for the spin components within the $ac$ plane ($J_i^{ac}$).
Considering the classical expression for the pitch angle along the zigzag chain, $\phi_c$\,=\,arccos($J_1$/4$J_2$), and the fact that neighboring magnetic ions lie in the same unit cell, i.e., $\phi_c$\,=\,$\pi k_c$ \cite{Pregelj} ($k_c$ is the $c$ component of {\bf k}), the comparison of {\bf k}\,+\,$\Delta${\bf k} and {\bf k} at 2.5\,K yields $(J_1/J_2)^{b}$\,$\sim$\,0.8$(J_1/J_2)^{ac}$. 
Such a scenario implies that $\Gamma_1$ and $\Gamma_2$ are subjected to different $J_1/J_2$, which leads to the difference {\bf $\Delta$k} between the corresponding modulations.
On the other hand, the very weak reflections at {\bf k}\,$-$\,$\Delta${\bf k} show an additional subtle 
($\sim$1\%) 
modulation of the $b$ component that is a direct consequence of the coupling between the two order parameters in Eq.\,(\ref{MF1}), trying to enforce the same size of the magnetic moments.
Our results thus demonstrate that the stripe-forming mechanism in frustrated spin-chains cannot rely solely on the frustrated interchain interactions, but must also involve magnetic exchange anisotropy.

In summary, we performed a detailed investigation of the magnetic orders in the VC, SDW and spin-stripe phases of $\beta$-TeVO$_4$.
Our results reveal that the intriguing spin-stripe phase develops as a superposition of two orthogonal SDW states that have slightly different modulation periods.
We attribute this to exchange anisotropy, leading to different effective interactions for the $b$ and $ac$ components of the magnetic moments. 
This constitutes a novel, anisotropy-driven, stripe-forming mechanism that may be active in numerous frustrated spin-chain systems, where sizable exchange anisotropy is regularly encountered.

\begin{acknowledgments}
We acknowledge the financial support of the Slovenian Research Agency (projects Z1-5443 and BI-HR/14-15-003 and program No. P1-0125) and the Swiss National Science Foundation (project SCOPES IZ73Z0\_152734/1).
This research has received funding from the European Union’s Seventh Framework Programme for research, technological development and demonstration under the NMI3-II Grant No. 283883.
The neutron diffraction experiments were performed at the Swiss spallation neutron source SINQ, at the Paul Scherrer Institute, Villigen, Switzerland, and at the reactor of the Institute Laue-Langevin, Grenoble, France.
M.H. acknowledges support of the Croatian Ministry of Science, Education and Sports and funding of the Croatian Science Foundation (project UIP-2014-09-9775).
\end{acknowledgments}

\section{Supplementary material}

\subsection{Derivation of the magnetic susceptibility tensor}\label{appA}
When the induced magnetization is linear in the applied magnetic field, $\mathbf{M} = \bm{\chi} \cdot \mathbf{H}$, as in paramagnets and antiferromagnets in low magnetic fields, magnetic torque ($\bm{\tau} = V \: \mathbf{M} \times \mathbf{H}$) directly probes the anisotropy of the magnetic susceptibility tensor $\bm{\chi}$. The tensor $\bm{\chi}$ is most often expressed in the coordinate system spanned by the crystal axes and can have nonzero off-diagonal elements, subjected to symmetry constrains. \\
\indent For the monoclinic system such as $\beta$-TeVO$_4$, when the susceptibility tensor is expressed in coordinate system spanned by the crystal axes $(a^*, b, c)$, the only off-diagonal element that symmetry allows is $\chi_{a^*c}=\chi_{c a^*}$ \cite{Newnham}. This implies that the $b$ axis is always the magnetic eigenaxis for a monoclinic system unless spontaneous symmetry breaking occurs, as it does in $\beta$-TeVO$_4$ below $T_{N1}$. Let us write the full susceptibility tensor explicitly as
\begin{equation}\label{eq:tensorchi}
\bm{\chi} =\begin{bmatrix} \chi_{a^*a^*}& \chi_{a^*b} & \chi_{a^*c}\\
\chi_{a^*b}& \chi_{bb} & \chi_{bc}\\
\chi_{a^*c} & \chi_{bc} & \chi_{cc}
\end{bmatrix}
\end{equation}
In our experiments the applied magnetic field was rotated in a chosen crystal plane and only the component of torque perpendicular to that plane was measured. Specifically, for the $a^*b$ plane the rotating magnetic field can be written as $\bm{H}= (\sin\theta, -\cos \theta, 0)$, where $\theta$ represents the angle the magnetic field direction makes with the $-b$ direction. Using this and the from of the susceptibility tensor (\ref{eq:tensorchi}), we obtain the expression for the measured component of the magnetic torque
\begin{equation}\label{eq:tauc}
	\tau_c= \dfrac{m}{2M_{mol}}H^2 [(\chi_{bb} - \chi_{a^*a^*})\sin 2\theta + 2\chi_{a^*b} \cos 2\theta]
\end{equation}
This expression can be simplified to 
\begin{equation}\label{eq:taucmeas}
	\tau_c = \dfrac{m}{2M_{mol}}H^2 \Delta \chi_{ba^*} \sin (2\theta-2\theta_0)
\end{equation}
where $\Delta \chi_{ba^*}$ represents the susceptibility anisotropy in the $a^*b$ plane, i.e. the difference between maximal and minimal anisotropy in that plane, and
\begin{align}
	\chi_{bb} &- \chi_{a^*a^*} =\Delta \chi_{ba^*} \cos \theta_0\\
	&\chi_{a^*b} =-1/2 \Delta \chi_{ba^*} \sin \theta_0
\end{align}
A rotation of the magnetic axes is observed as a change of $\theta_0$.
In the magnetic torque experiments both $\Delta \chi$ and $\theta_0$ can be measured simultaneously. 
In this way, the off-diagonal tensor component can be obtained along with the difference of the two diagonal components in the measured plane.
When the measurements are repeated in two other planes and the susceptibility along at least one of the axes is independently measured, the temperature dependence of the full susceptibility tensor can be obtained. 
Finally, the susceptibility eigenvalues are extracted by diagonalizing the susceptibility tensor.

\subsection{Details of the magnetic structure}\label{appB}

The magnetic structure model dictates the magnetic moment at a particular V site to follow an elliptical helix with pitch along the magnetic $\bf{k}$ vector,
\small 
\begin{equation}
{\bf{S}}_{n}({\bf{r}}_{i})  =  {\bf{S}}_{0\,n}^{\text{Re}}\cos({\bf{k}}\cdot{\bf{r}}_{i} - \psi_{n}) 
 +  {\bf{S}}_{0\,n}^{\text{Im}}\sin( {\bf{k}}\cdot{\bf{r}}_{i} - \psi_{n}). 
\end{equation} 
\normalsize
Here, the vector ${\bf{r}}_i$ defines the origin of the $i$-th cell and $n$=1-4 denotes the four V positions within the crystallographic unit cell.
The complex vector ${\bf{S}}_{0\,n}$ is determined by its real and imaginary components, ${\bf{S}}_{0\,n}^{\text{Re}}$ and ${\bf{S}}_{0\,n}^{\text{Im}}$, defining the amplitude and the orientation of the magnetic moments, while $\psi_{mn}$ denotes the phase shift.
We assume the same moment ${\bf{S}}_{0\,j}$\,$\equiv$\,${\bf{S}}_{0\,j+1}$ for $j$\,=\,1,3.
In this respect, the parameters for the spin-density-wave (SDW) and vector chiral (VC) states are given in Table\,\ref{mag-struc-model}.
Finally, we point out that in the SDW state, the magnetic structure is amplitude modulated so that ${\bf{S}}_{0\,n}^{\text{Im}}$\,$\equiv$\,0, which are thus omitted in the corresponding table.

\begin{table} [!]
\caption{Parameters of the best magnetic structure model  for the SDW and VC phases at 3.5\,K and 1.7 K, respectively, for two independent magnetic atoms (V$_1$ and V$_3$), and magnetic phases $\psi_{n}$ for all of the magnetic V$_{n}$ atoms in the unit cell ($n$\,=\,1-4). The sites V$_{2}$ and V$_{4}$ are obtained from V$_{1}$ [$0.7042(1)$, $0.1757(1)$, $0.6416(1)$] and V$_{3}$ [$0.2958(1)$, $0.6757(1)$, $0.8584(1)$], respectively, by symmetry element $c(x,1/4,z)$. The orientation of the moments is given in the $a^*bc$ coordinate system. 
\label{mag-struc-model}}
\begin{ruledtabular}
\begin{tabular}{c|c|c||c|c}
\multicolumn{5}{c}{The SDW phase at 3.5\,K}\\
\hline
     $s$ = Re, Im   & V$_1^{\text{Re}}$ & V$_3^{\text{Re}}$ & $n$ &  $\psi_{n}$ \\
\hline
$S_{0\,x}^s$ /$|S_{0}^s|$        & 0.35(5)   &   0.00(5) & 1 &   0.41(5) \\
$S_{0\,y}^s$ /$|S_{0}^s|$        & 0.00(5)   &   0.00(5) & 2 &  0.62(5) \\
$S_{0\,z}^s$ /$|S_{0}^s|$        & 0.00(5)   &   0.74(5) & 3 &    0.00(5) \\ \cline{1-3}
$|{\bf S}_{0\,m}^s|/|{\bf S}_{0}|$  & 0.35(5)  &  0.74(5)  & 4 &    0.21(5) \\
\end{tabular}
\end{ruledtabular}
\\
\begin{ruledtabular}
\begin{tabular}{c|c|c|c|c||c|c}
\multicolumn{7}{c}{The VC phase at 1.7\,K}\\
\hline
     $s$ = Re, Im   & V$_1^{\text{Re}}$ & V$_1^{\text{Im}}$ & V$_3^{\text{Re}}$ & V$_3^{\text{Im}}$ & $n$ &  $\psi_{n}$ \\
\hline
$S_{0\,x}^s$ /$|S_{0}^s|$        & 0.45(5)   & 0.70(5)   &   0.00(5) &   0.04(5) & 1 &   0.00(5) \\
$S_{0\,y}^s$ /$|S_{0}^s|$        & 0.13(5)   & 0.39(5)   &   0.68(5) &   0.00(5) & 2 &  0.91(5) \\
$S_{0\,z}^s$ /$|S_{0}^s|$        & 0.00(5)   & 0.00(5)   &   0.00(5) &   0.97(5) & 3 &    0.78(5) \\ \cline{1-5}
$|{\bf S}_{0\,m}^s|/|{\bf S}_{0}|$  & 0.47(5)   & 0.80(6)   &  0.68(5)  &  0.97(5)  & 4 &    0.50(5) \\
\end{tabular}
\end{ruledtabular}

\end{table}

\subsection{Temperature dependence of the incommensurate phases}\label{appC}

The description of the temperature evolution of the incommensurate phases within a single model relies on the Landau free energy \cite{Lawes}
\small 
\begin{equation}
F = a_1 (T-T_{N1})\sigma_1^2 + a_2 (T-T_{N2})\sigma_2^2 + V + O(\sigma^4).
\end{equation} 
\normalsize
Here $a_1$ and $a_2$ are constants related to the magnetic order parameters $\sigma_1$ and $\sigma_2$, $V$ is the coupling term between $\sigma_1$ and $\sigma_2$ explained in the main text, and $O(\sigma^4)$ stands for the rest of the fourth order terms that are beyond the scope of our study.
The above expression describes a typical situation in spiral multiferroics \cite{Wang}.
Namely, the emergence of the incommensurate magnetic order at $T_{N1}$, due to establishment of $\sigma_1$, and its change at $T_{N2}$, where also $\sigma_2$ is established.
We note that the VC phase in $\beta$-TeVO$_4$ phase is established only at $T_{N3}$, when the magnetic wave vectors corresponding to  $\sigma_1$ and $\sigma_2$ become equal.

\end{document}